\documentclass[amssymb,twocolumn,aps,prb]{revtex4-2}

\usepackage{graphicx}
\usepackage{dcolumn}
\usepackage{bm}
\usepackage{mathtools}
\usepackage{float}
\usepackage[T1]{fontenc}

\DeclareUnicodeCharacter{2212}{\textendash}

\begin{document}

\preprint{APS/123-QED}

\title{Measurement of exciton fraction of microcavity exciton-polaritons using transfer-matrix modeling}

%
%


\author{
J. Beaumariage,$^1$
Z. Sun,$^{1,2}$
H. Alnatah,$^1$
Q. Yao,$^{1,3}$
D. M. Myers,$^{1,4}$
M. Steger,$^1$
K. West,$^5$
K. Baldwin,$^5$
L. N. Pfeiffer,$^5$
M. C. A. Tam,$^6$
Z. R. Wasilewski,$^6$
and D. W. Snoke$^1$
}

\affiliation{$^1$Department of Physics and Astronomy, University of Pittsburgh, 3941 O'Hara St., Pittsburgh, PA 15260, USA\\
 $^2$Current Address: State Key Laboratory of Precision Spectroscopy, East China Normal University, Shanghai 200241, China\\
 $^3$Current Address: Joint Quantum Institute, University of Maryland and National Institute of Standards and Technology, College Park, Md 20742, USA.\\
 $^4$Current Address: Kulicke \& Soffa Industries, Inc., 1005 Virginia Dr, Fort Washington, PA 19034\\
 $^5$Department of Electrical Engineering, Princeton University, Princeton, NJ 08544, USA\\
 $^6$Department of Electrical and Computer Engineering, University of Waterloo, Waterloo, Ontario N2L 3G1, Canada\\
}

\date{\today}

\begin{abstract}
We present a careful calibration of the exciton fraction of polaritons in high-$Q$ ($\sim 300,000$), long-lifetime ($\sim 300$ ps), GaAs/AlGaAs microcavities.  This is a crucial parameter for many-body theories which include the polariton-polariton interactions. It is much harder to establish this number in high-$Q$ structures compared to low-$Q$ structures, because the upper polariton is nearly invisible in high-$Q$ cavities.  We present a combination of photoluminescence, photoluminescence excitation, and reflectivity measurements to highly constrain the fit model, and compare the results of this model to the results from low-$Q$ structures. We present a fitted curve of exciton fraction as a function of the lower polariton energy for multiple samples which have been used in prior experiments.

\end{abstract}


\maketitle

\section{\label{sec:intro} Introduction}

The GaAs/AlGaAs microcavity structure for exciton-polaritons is a well explored system, in which a wide variety of physical phenomena have been observed, including Bose-Einstein condensation \cite{snoke_science_2007} , propagation lengths of several millimeters with lifetimes of ~180 ps \cite{steger_optica_2015}, and many other interesting effects \cite{mukherjee_dec2019,yao_dec2022,myers_dec2018,
deng_polariton_lasing_vs_photon_lasing,
deng_quantum_degenerate_polaritons_equilibrium,
bloch_high_temp_ultrafast_parametric_amplification,
Sanvitto2010_persistent_currents_quantized_vortices_superfluid,
Baumberg_inversion_hysteresis_spin_condensate,
savvidis_angle_resonant_stimulated_polariton_amplifier,
lagoudakis_p_next_nearest_neighbor_coupling_spinor_condensates,
lagoudakis_k_spontaneous_pattern_formation_condensate,
Ostovskaya_topological_phase_transition,
assman_optical_vortex_switching}. The exciton fraction of the polaritons in these experiments is essential in various calculations such as polariton-polariton interaction strength \cite{yoseob_jun2017,snoke_reanalysis_2023} as well as getting accurate measurements of the absolute polariton density from photoluminescence (PL) \cite{myers_dec2018, mitprl,hassan_criticial_flucturations,Hassan_coherence_measurements_power_law_2d_condensates}. In a short-lifetime (low-$Q$) microcavity, calculating the exciton fraction is relatively straightforward, typically using a Rabi model \cite{balili_thesis_2009}. In those samples, both the upper and lower polaritons may be observed in PL images as well as reflectivity images for such samples, and so multiple approaches are available for calculating the exciton fraction. However, in long-lifetime samples, neither polariton branch may be observed in the reflectivity spectra, because the polariton lines are so spectrally narrow that they will not appear without ultra-high spectral resolution. Furthermore, the upper polariton's PL is so weak as to be unobservable in many regions of interest. A simple way to understand why the upper polariton PL is so weak is to think in terms of the branching ratio for decay processes of the upper polaritons. In low-Q samples, emission of PL photons from the upper polariton states occurs at about the same rate as emission of phonons to jump down into a lower polariton state, but in high-Q samples, jumping down into lower polariton states occurs much faster than photon emission from the upper polaritons.

This makes determination of the exciton fraction extremely difficult. In this paper we outline how to measure the upper polariton through photoluminescence excitation (PLE) spectroscopy, and how to combine that information with a transfer-matrix simulation of the samples \cite{beaumariage_tmm_matlab} in order to calculate the exciton fraction.




The structures used in our experiments are created by putting two highly reflective distributed Bragg reflectors (DBRs) in close proximity with each other in order to form a standing wave of light with 3 antinodes.  At the antinodes of the standing wave we have a series of quantum wells. This design results in large electric fields interacting with the excitons inside the quantum wells. The entire sample is grown through molecular-beam epitaxy (MBE).

The simplest model for this system is a two-level system \cite{deng_review_2010} in which the excitons are approximated as having an infinite mass and thus a single energy $E_x$ which does not depend on in-plane momentum $k_{\parallel}$. The cavity photon, however, does have a $k_{\parallel}$-dependent energy given by

\begin{equation}\label{eqn:E_cavity}
  E_{c} (k_{\parallel}) = \sqrt{E_{c0}^2 + \bigg(\frac{\hbar c k_{\parallel}}{n_c}\bigg)^2},
\end{equation}

where $E_{c0}$ is the cavity energy at zero in-plane momentum due to the standing wave between the two DBRs, and $n_c$ is the index of the cavity. These two modes couple with a strength strength $g$, resulting in the Hamiltonian

\begin{equation}\label{eqn:starting_hamiltonian}
  \hat{H} = \sum_{k_{\parallel}} E_x \hat{a}^\dagger _{k_\parallel} \hat{a} _{k_\parallel} + \sum_{k_{\parallel}} E_c(k_{\parallel}) \hat{b}^\dagger _{k_\parallel} \hat{b} _{k_\parallel} + \sum_{k_{\parallel}} g(\hat{a}^\dagger _{k_\parallel} \hat{b} _{k_\parallel} + \hat{a}_{k_\parallel} \hat{b}^\dagger  _{k_\parallel}).
\end{equation}

This Hamiltonian may be diagonalized to find the dispersion relationships of the new eigenstates (the polaritons):

\begin{equation}\label{eqn:E_polaritons}
E_{UP,LP} (k_\parallel) = \frac{1}{2} \left( E_x + E_c(k_\parallel) \pm \sqrt{4g^2 + \big (E_c(k_\parallel) -E_x  \big )^2} \right),
\end{equation}

where the lower polariton states $| L \rangle$ may be written in terms of the exciton states $| X \rangle$ and the cavity states $| C \rangle$ as

\begin{equation}\label{eqn:LP_states}
| L_{k_{\parallel}} \rangle = X_{k_{\parallel}} | X_{k_{\parallel}} \rangle + C_{k_{\parallel}} | C_{k_{\parallel}} \rangle.
\end{equation}

The quantity $E_c(k_\parallel) -E_x$ appears frequently in this model and is called the ``detuning,'' that is, the energy of the bare cavity photon relative to the bare exciton. This parameter is typically controllable experimentally by a wedge of the optical cavity that causes the photon energy of the cavity mode to vary continuously across the MBE-grown wafer, while the exciton energy is relatively constant for the quantum wells.

The quantity $|X_{k_{\parallel}}|^2$ is the exciton fraction of the lower polaritons at the specified in-plane momentum. As the lower polaritons are frequently the polaritons of interest, $|X_{k_{\parallel}}|^2$ is colloquially referred to as just the exciton fraction. Additionally, we are frequently interested in the exciton fraction at zero in-plane momentum, which we will denote simply as $|X|^2$. The cavity fraction of the lower polaritons $|C_{k_{\parallel}}|^2$ may easily be calculated from the exciton fraction as $|C_{k_{\parallel}}|^2 = 1 - |X_{k_{\parallel}}|^2$, and so we focus our efforts on finding the exciton fraction. During the diagonalization procedure the exciton fraction is found to be
\begin{equation}\label{eqn:ex_frac_2state}
|X_{k_{\parallel}}|^2 = \frac{1}{2} \left( 1 - \frac{ E_x - E_c(k_\parallel)  }{\sqrt{4g^2 + \big (E_c(k_\parallel) -E_x  \big )^2}} \right).
\end{equation}

In practice, this equation is not so useful for direct calculation. In our long-lifetime (high-Q factor) samples, neither the exciton nor the cavity mode are visible in photoluminescence (PL) measurements. Additionally, they are not visible in reflectivity measurements. However, this equation can still give us important insights. In particular, we note that the derivative of the lower polariton energy in Equation (\ref{eqn:E_polaritons}) with respect to the exciton energy is exactly equal to the exciton fraction  (\ref{eqn:ex_frac_2state}). We take this to be a sensible definition for the exciton fraction of polaritons in our more complicated model outlined in Section \ref{sec:TMM_fine}.

\begin{equation}\label{eqn:ex_frac_general}
|X_{k_{\parallel}}|^2 = \frac{d}{dE_x} E_{lp}(k_\parallel)
\end{equation}

In addition to being unable to directly measure the exciton and cavity photon energies, we find that the upper polariton is not present in PL or reflectivity measurements of our long-lifetime samples. Therefore, of the many quantities used in this two-state Rabi model, only the lower polariton's energy may be easily measured. In the paper we will discuss what measurements are possible, how to perform them, and how to use the various measurements to arrive at an exciton fraction. 

In Section \ref{sec:ks_pl_refl} we will discuss how to calibrate an angle-resolved imaging setup and take PL measurements of the lower polariton. Then in Section \ref{sec:PLE} we will discuss how to measure the upper polariton at normal incidence through photoluminescence excitation (PLE), including the necessary dynamic background subtraction. In Section \ref{sec:TMM_broad} we will discuss how to use our calibrated optical setup to take reflectivity measurements and how to create a transfer-matrix method (TMM) simulation of the sample, including fitting it to the measured reflectivity. In Section \ref{sec:TMM_fine} we will discuss further fitting our TMM simulation to the PL and PLE data. Specifically we will demonstrate all these techniques on data taken from a specific sample used in many prior experiments (Sample 4-6-15.1) at a location on the wafer where the detuning is near zero, that is, the exciton and photon states at $k_{\parallel}=0$ are nearly resonant. Then we will show the results of performing this fit at multiple locations on the sample, and a two parameter fit relating lower polariton energy to exciton fraction. We also include a table giving the fits for many samples we have worked with over the years. Last, in section \ref{sec:low_Q} we will perform this process on a low-Q sample along with several other measurements which act as a control study to confirm our process is sound, and discuss the uses and limitations of our approach.

\section{\label{sec:ks_pl_refl} Angle Calibration for PL and Reflectivity}

\begin{figure*}
  \includegraphics[width= 1 \linewidth]{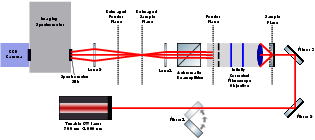}
  \caption{Illustration showing a basic Fourier imaging setup. Mirror 1 may be flipped into the laser path to switch to a reflection geometry, or left out of the path to create a transmission geometry. Due to the opaque substrate of our samples, a reflection geometry is utilized for collecting PL, reflectivity, and PLE data. However, the transmission geometry is utilized along with a transmission grating in order to collect the calibration data shown in Figure \ref{fig:angle_calibration}. Lens 2 may be chosen and positioned to image either the real-space sample plane or the Fourier plane onto the slit of the spectrometer. Additional optics such as longpass filters and polarizers may be placed after Lens 1.} \label{fig:optics_setup}
\end{figure*}

\begin{figure}
  \includegraphics[width= 1 \linewidth]{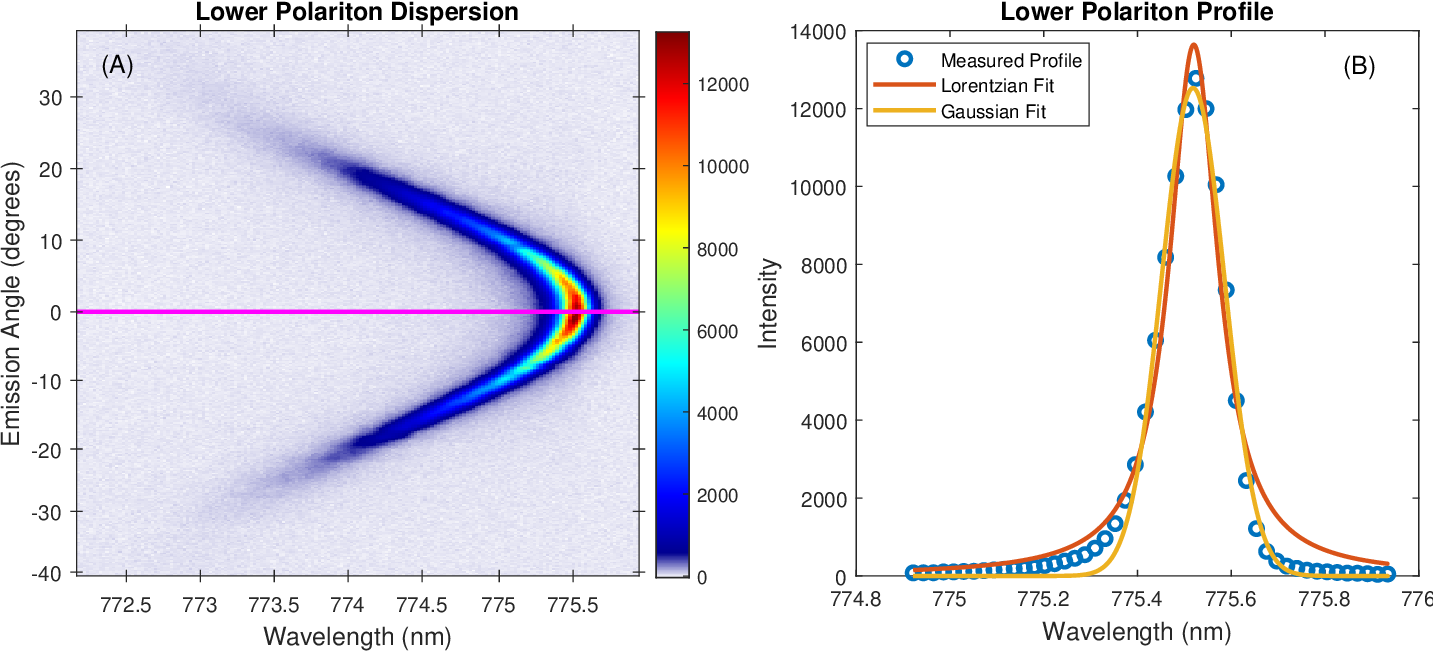}
  \caption{Image A: PL image of the lower polariton created by non-resonant pumping with a laser tuned to approximately 725 nm, which corresponds to a local minimum in the reflectivity as shown in figure \ref{fig:refl_fit}. This PL has very low noise, and so it is used to create the mask image in Figure \ref{fig:PLE_background}. Image B: The intensity profile of the PL at emission angle of zero degrees. We see that neither Lorentzian nor Gaussian fits completely capture the shape of our line. However, both fits are suitable for extracting a line center and linewidth. We use the Lorentzian fit to extract the line center at every angle to generate our dispersion curves.}\label{fig:LP_PL_profile}
\end{figure}

The main decay channel for lower polaritons is to produce a photon exiting the sample. In this process, energy and in-plane momentum are conserved. Thus by observing the PL of the lower polaritons we may directly measure their dispersion relationship. In order to do this, we require an angle-resolved imaging setup, also known as a Fourier imaging setup. Such a setup is shown in Figure \ref{fig:optics_setup}. At the Fourier plane of the microscope objective, a Fourier-transformed image of the sample plane is produced. This optical Fourier transform takes a real space image and produces a $k$-space image. That $k$-space image is then re-imaged onto the entrance slit of an imaging spectrometer. By closing the slit down and setting the spectrometer to an appropriate wavelength setting, we can directly produce images such as the one shown in Figure \ref{fig:LP_PL_profile}A.

It is essential to calibrate the vertical and horizontal axes so that we may associate a $k_\parallel$ value with each energy. However, the spectrometer produces images in which the wavelength varies approximately linearly with the horizontal pixel number. Furthermore, our optics refract light rays based on their angle rather than their in-plane momentum. Therefore, although the theory of polaritons is best expressed in terms of energy and momentum, the measured data is best shown in terms of wavelength and angle. For most of this paper we will be working in these experimentalist's units, and we will convert back to energy and momentum at the end.

\begin{figure}
  \includegraphics[width= 1 \linewidth]{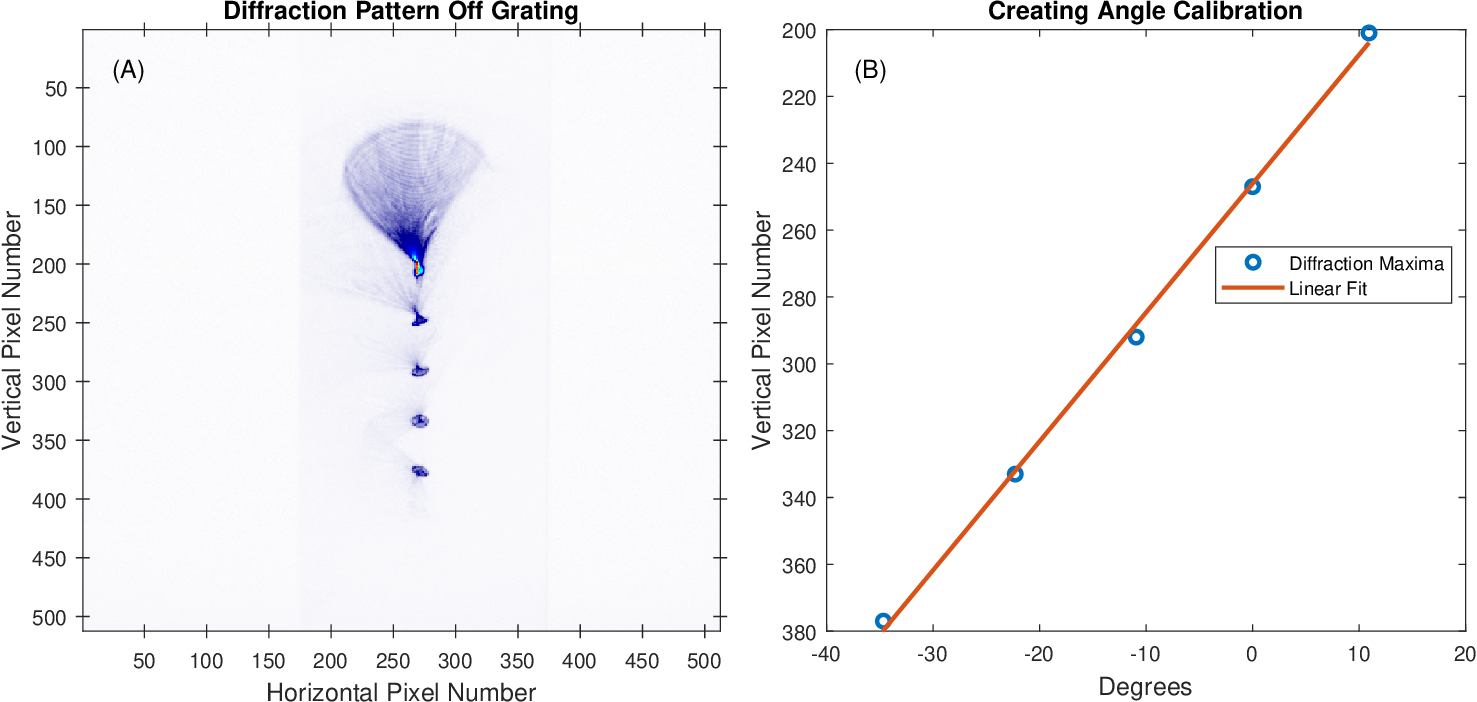}
  \caption{Data used to create the calibration of our angle resolved imaging. Image A is data collected using the transmission setup with a transmission grating as shown in figure \ref{fig:optics_setup}. Peak locations are extracted and plotted as shown in image B. These data points are then fit with a linear equation in order to create a calibration between vertical pixel number on our camera and the emission angle of the light from our samples.}\label{fig:angle_calibration}
\end{figure}

In order to create a calibration relating vertical pixel number on our camera to emission angle of light rays in our setup, we used the setup illustrated in Figure \ref{fig:optics_setup}. In the sample plane we placed a GT13-03 visible transmission grating (300 Groves/mm) from Thorlabs. This produces the image shown in Figure \ref{fig:angle_calibration}A. 

The laser was a Solstis laser from M Squared (tunable Ti Sapphire laser with narrow linewidth and stable power). We set the laser to a wavelength of 775 nm. Through the grating equation we are able to identify each of the dots of laser light in the image as a known angle. The laser dot's measured pixel numbers versus the calculated angles are shown in Figure \ref{fig:angle_calibration}B. This fit is then applied to the entire camera's vertical axis for future images. In principle we could do this calibration at many laser wavelengths in order to get the calibration as a function of wavelength. However, we found the wavelength correction to the calibration to be relatively small over our wavelengths of interest, and so we have ignored it for simplicity.

The calibration of our horizontal axis was comparatively straight forward. The Solstis laser is equipped with a WS/6 Wavelength Meter from HighFinesse which measures its wavelength. We send the laser directly into the spectrometer by flipping mirror 1 into our laser's path, then we place a silver mirror in the sample plane of the microscope objective. Next we tune the laser across a variety of wavelengths that span the entire screen, taking an image at each wavelength. Using this data we can create a calibration relating wavelength to horizontal pixel number. Conveniently, this method of calibration means that our laser wavelength measurements and our image wavelength measurements are consistent with each other.

With our angle-resolved imaging setup fully calibrated, we can take PL measurements of the lower polariton. Figure \ref{fig:LP_PL_profile}A shows the dispersion relationship of the lower polaritons created by pumping the sample non-resonantly with a continuous wave (CW) laser tuned to ~725 nm. A line is drawn along the data corresponding to light at normal incidence. The PL profile along that line is shown in Figure \ref{fig:LP_PL_profile}B, with corresponding fits using Gaussian and Lorentzian distributions. We find that neither distribution quite fits all aspects of the line shape; this is true whether we fit the distribution in energy or wavelength. Using the Lorentzian fit, we perform this procedure at every row of the image to extract the line center at each angle $\lambda(\theta)_lp,pl$. We keep as much of the data as we can, but do trim off the higher angle data as the signal-to-noise ratio is too low to get a good fit. We fit the data in the $\pm 10$ degree range with a simple parabola in order to find the true zero of the angle axis. We then shift where the $\theta = 0$ line is on our image by shifting the entire angle axis up or down. This is necessary as small shifts of the angle axis happen in the day-to-day alignments of the system (particularly when inserting and aligning new samples).

It is at this point easy enough to convert the data extracted from images like Figure \ref{fig:LP_PL_profile}A to energy and in-plane momentum. In principle, these data could be fit with Equation (\ref{eqn:E_polaritons}) even without the upper polaritons present in our data. However, in practice, this is a very unconstrained fit. To test this first we used Equation (\ref{eqn:E_polaritons}) to generate theoretical data for polaritons which are 50\% excitonic at $\theta = 0$ over a range of $\pm$30 degrees. Then fitting Equation (\ref{eqn:E_polaritons}) to that ideal generated data, it is possible to find extremely good fits that are anywhere between 10\% excitonic to 75 \% excitonic by varying the fitting parameters; a gigantic and useless range. If we try fitting the true experimental data extracted from Figure \ref{fig:LP_PL_profile} we find nearly identical results. By varying the fitting parameters, we can obtain fits that are nearly identical quality with an exciton fraction anywhere from 10\%  to 75 \%.

Thus, we can conclude that simply fitting the dispersion curve of the lower polaritons with Equation (\ref{eqn:E_polaritons}) is not sufficient to precisely determine an exciton fraction. We need to perform additional measurements to constrain our fitting.

\section{\label{sec:PLE} Measurement of the Upper Polariton via PLE}

The main difference between our long-lifetime samples and short-lifetime samples is the number of periods composing the top and bottom DBRs. Our older short-lifetime samples had 20 periods in the bottom DBR and 16 periods in the top DBR, whereas our modern long-lifetime samples have 40 periods in the bottom DBR and 32 periods in the top DBR. The greatly improved reflectivity of the DBRs is responsible for the long lifetime of our polaritons. However, this also means that the polaritons have very narrow linewidths. The linewidths are so narrow that we are not able to observe either polariton branch when we measure the angle-resolved reflectivity such as shown in Figure \ref{fig:refl_fit}A. Furthermore, the PL of the upper polaritons is nearly impossible to see because the upper polaritons efficiently scatter down into lower polaritons through phonon emission. (In short-lifetime samples, much more of the PL is visible because the branching ratio of PL versus phonon down-conversion favors the PL emission, which indirectly implies that the phonon emission time for down conversion to lower polaritons is of the order of a few picoseconds.) The decay of upper polaritons is dominated by this down-conversion process, unlike the lower polaritons, which predominantly decay through photon emission.

While this means that upper polaritons are not observable in PL, it also means the upper polaritons are well suited for measurement through photoluminescence excitation (PLE). The basic idea of PLE is to sweep the wavelength of the pump laser while observing the total intensity of the lower polaritons. When the laser is resonantly tuned to the upper polaritons, maximal absorption should occur, and then the upper polaritons will scatter down into lower polaritons. It is worth pointing out that a direct absorption measurement utilizing a transmission geometry is possible using the setup shown in Figure \ref{fig:angle_calibration}. However, our substrate of choice, GaAs is not transparent at the wavelengths of interest.

\begin{figure}
  \includegraphics[width= 1 \linewidth]{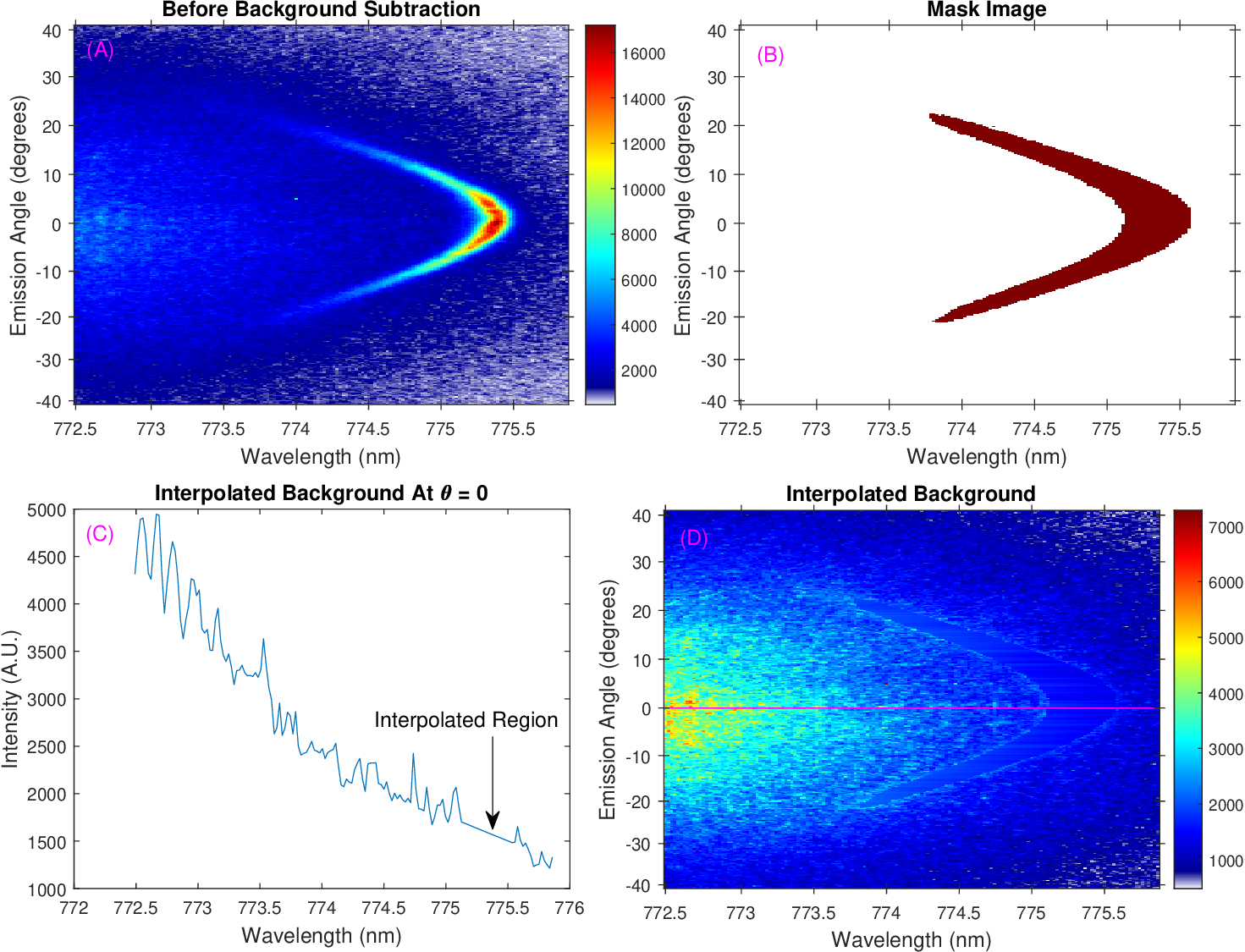}
  \caption{These images show the background subtraction method for our PLE sweeps. Image A is a raw image taken while sweeping the pump laser across the upper polariton's resonance. We see a large amount of noise compared to the non-resonant pumping shown in figure \ref{fig:LP_PL_profile}A. For this reason Figure \ref{fig:LP_PL_profile}A is used to create the mask image shown here in image B. The pixels in image A which correspond to white pixels in image B are taken to be purely background noise. The noise pixels are then fit with a third order polynomial to interpolate the average value of the background inside the red pixels. This interpolated result  at normal incidence is shown in image C. The process is done at each angle independently to produce the full background image shown in image D. Image D may then be subtracted from image A in order to arrive at image \ref{fig:PLE_results}B.}\label{fig:PLE_background}
\end{figure}

The biggest challenge we face in our PLE measurements is that because the upper polariton's wavelength is only about 10 nm away from the lower polariton, the noise of our laser scattering can easily overwhelm the signal of our measurement. To avoid this, we first use basic optical methods to filter out scattered laser light. The laser in our setup is vertically polarized, and the lower polaritons produced through the phonon emission are unpolarized. Therefore, we can put a horizontal polarizer into our imaging setup to remove a large amount of the scattered laser light while only removing half of our PL signal. Furthermore, we have a tunable longpass filter from Semrock, which we put into our imaging setup. This filter is tuned by simply rotating the angle of the filter slightly. We tune it so that most of the lower polariton PL may pass, while most of the laser is filtered out. Then, without changing any optics, we sweep the wavelength of our laser while taking a series of images. One such image is shown in Figure \ref{fig:PLE_background}A.

As we can see when comparing this image to Figure \ref{fig:LP_PL_profile}A, a large amount of laser noise is still present in the image. Worse yet, as we sweep the wavelength of the pump laser, the noise in the image shifts along the wavelength axis. The majority of this noise comes from laser light reflected off the sample, which means it changes with position on the sample. Therefore, there is no simple way to just take a background image, and subtract off the noise. Instead, we devised a method to interpolate all the pixels which do not contain PL to give the value of the background in the pixels which do contain PL.

We start by using the clean image of the PL created through non-resonant pumping shown in Figure \ref{fig:LP_PL_profile}A to determine which pixels contain PL signal. That information will then carry over to Figure \ref{fig:PLE_background}A, telling us which pixels contain both PL signal and background versus which pixels contain only background. We will call the intensity of the clean image in Figure \ref{fig:LP_PL_profile}A $P_c(\lambda,\theta)$ and the intensity of the noisy image in Figure \ref{fig:PLE_background}A $P_n(\lambda,\theta)$.

We define a threshold intensity $T$, such that pixels in figure \ref{fig:LP_PL_profile}A with intensities above this threshold are considered to have polariton PL in them, while pixels with intensities below this threshold are assumed to contain only background and no PL. The value of the threshold is determined by experience using our data processing interface \cite{beaumariage_tmm_matlab}, which allows the user to control this choice by producing the mask image shown in \ref{fig:PLE_background}B. A threshold value that is lower than needed is not a problem, while a value too high can be a problem. Therefore, we erred towards lower threshold values, which is why the red pixels in the mask image seem larger than the PL in the clean image.  Mathematically we can write the intensity profile of the mask image as
\begin{align}
M(\lambda,\theta) &= \left\{\begin{aligned}
1 && P_c(\lambda,\theta) \geq T
\\
0 && P_c(\lambda,\theta) < T
\end{aligned}\right.\\
\end{align}

We refer to the pixels above the threshold as red pixels and pixels below the threshold as white pixels. Along each row of the image in Figure \ref{fig:PLE_background}A, we fit the white pixels with some curve, then use that curve to interpolate the value of the background within the red pixels. We find that a third order polynomial is sufficient to capture the intensity behavior of the white pixels. We denote this best fit as $Polyfit(M=0,\lambda,\theta)$, where $M=0$ indicates that the fit is created using only the white pixels. A separate fit is performed for each row of the data. We write our constructed background as
\begin{align}
B(\lambda,\theta) &= \left\{\begin{aligned}
P_n(\lambda,\theta) && M(\lambda,\theta) = 0
\\
Polyfit(M=0,\lambda,\theta) && M(\lambda,\theta) = 1
\end{aligned}\right.
\\
\end{align}

\begin{figure}
  \includegraphics[width= 1 \linewidth]{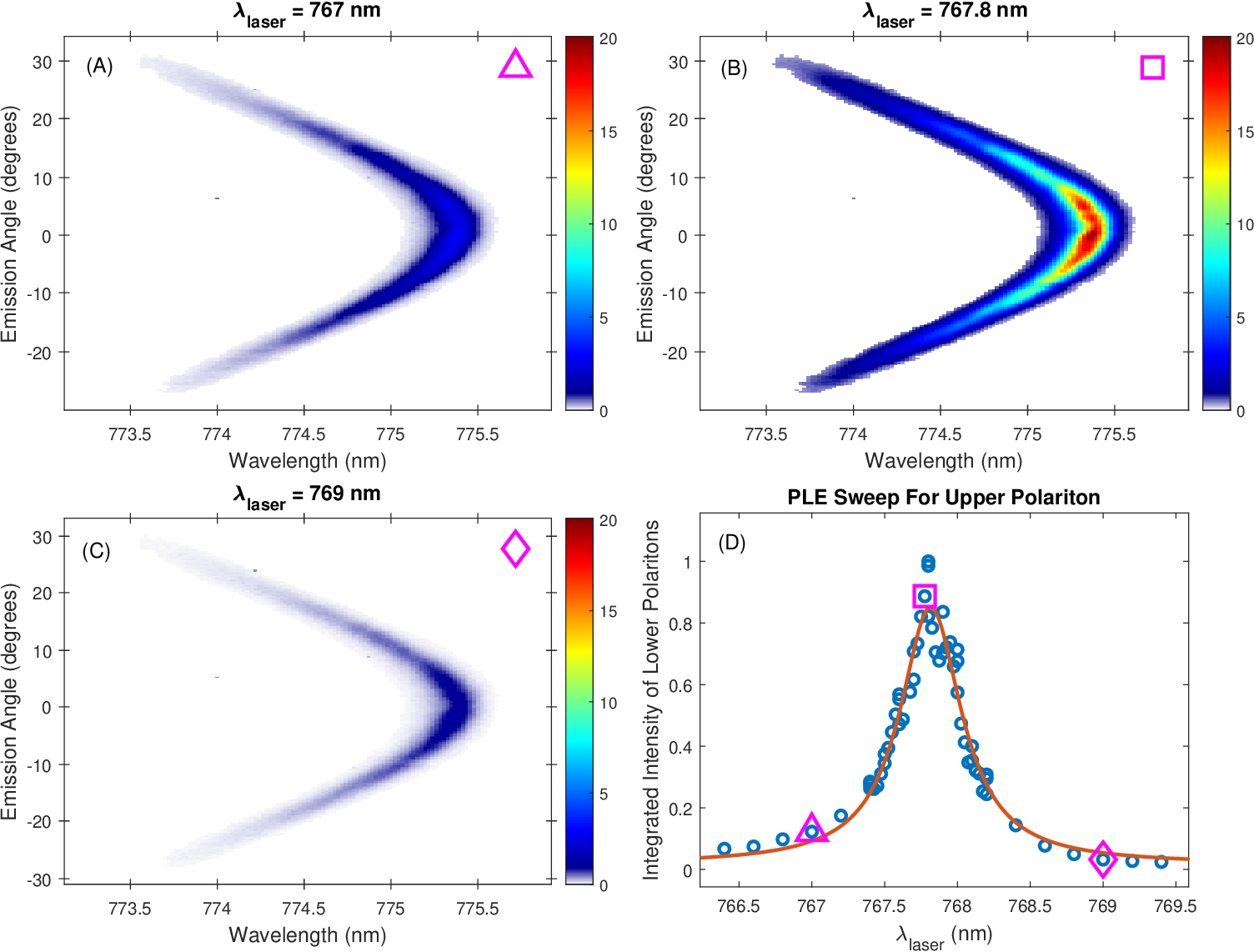}
  \caption{Images of the lower polariton as the pump laser wavelength is swept across the upper polariton's resonance. The background subtraction is handled as demonstrated in Figure \ref{fig:PLE_background}. Images A, B, and C are then integrated to produce the data points shown in image D. We see a clear peak form corresponding to the upper polariton's resonance. }\label{fig:PLE_results}
\end{figure}

The constructed background corresponding to the image in Figure \ref{fig:PLE_background}A is shown in figure \ref{fig:PLE_background}D, and the intensity profile along the line drawn at $\theta = 0$ is shown in Figure \ref{fig:PLE_background}C. The interpolated region is labeled and quite visible, as the random fluctuations seen in the real noise are not present in the interpolated region. However, the integrals that follow in the next steps reduce the impact of fluctuations. Our interpolation scheme has produced a background image which is sufficient for our purposes. We can now subtract off the background image to produce an image of just the lower polariton PL. We call this background subtracted image's intensity $P(\lambda,\theta)$, giving
\begin{equation}\label{eqn:cleaned_pl}
  P(\lambda,\theta) = P_n(\lambda,\theta) - B(\lambda,\theta).
\end{equation}
 This procedure is repeated at each wavelength in our laser sweep. Examples of these background subtracted images are shown in Figures \ref{fig:PLE_results}A,\ref{fig:PLE_results}B, and \ref{fig:PLE_results}C.

Now that we have produced images of lower polariton PL free of laser noise, we integrate the images to obtain the total intensity of lower polaritons as a function of the pump laser's wavelength. The images are actually two-dimensional slices of the full three-dimensional $k$-space. Because the dispersion relationship is radially symmetric in the in-plane momentum directions, we perform this integral using circular symmetry. Numerically, this integrated lower polariton intensity is
\begin{equation}\label{eqn:num_int_LP}
  I = \sum_{\lambda} \sum_{\theta} P(\lambda,\theta) \; 2 \pi \theta \; \Delta  \theta \; \Delta  \lambda,
\end{equation}
where $\Delta \theta$ and $\Delta \lambda$ are determined by the resolution of our setup.

In order to convert from counts on a camera to absolute polariton density, it is necessary to correct by the lifetime of the polaritons. This lifetime is both angle and wavelength dependent. In principle, if the temperature of the lower polaritons changed during the measurement, this could lead to a change of the calibration of the lower-polariton density. However, the process of sweeping the pump laser wavelength does not significantly change the distribution of the polaritons. This means that the distribution of $P(\lambda,\theta)$ does not change from image to image. Since our goal is not to determine absolute polariton populations, but rather only to see relative differences in this integrated intensity, we may ignore this correction in our measurement.

Thus, we have a series of images created by our laser sweeps, and we can plot the intensity as a function of the laser wavelength as shown in Figure \ref{fig:PLE_results}D. We see a clear peak formed in the center of our range, which is well fit by a Lorentzian distribution. Care was taken to increase the density of our data points in the most important spectral region. Three data points are selected and enclosed by a triangle, square, and a diamond. The background subtracted images corresponding to these three data points are shown in Figures \ref{fig:PLE_results}A, \ref{fig:PLE_results}B, and \ref{fig:PLE_results}C. One can easily see that Figure \ref{fig:PLE_results}B is noticeably brighter than the other two images. From our Lorentzian fit we extract the line center, and conclude this is the wavelength of the upper polariton $\lambda_{up,ple}$. Furthermore, because our pump laser was a cone centered on normal incidence ranging over $\pm$ 5 degrees, we conclude specifically we have measured the upper polariton at $k_{\parallel} = 0$. In Section \ref{sec:low_Q} we report these measurements on low-Q samples where the upper polariton is visible in reflectivity measurements in order to quantify the discrepancy between PLE measurements and reflectivity measurements.

\section{\label{sec:TMM_broad} Fitting a TMM Model to the Broad Reflectivity}

As discussed in Section \ref{sec:ks_pl_refl}, simply fitting the lower polariton PL with Equation (\ref{eqn:E_polaritons}) results in a very weakly constrained fit and a wide range of possible exciton fractions. The measurement of the upper polariton at $k_{\parallel} = 0$ acts as a constraint on our fitting parameters, and greatly improves the fits. As discussed above, without the PLE constraint, we found fits of the data in Figure \ref{fig:LP_PL_profile}A with exciton fractions ranging from 10\%  to 75 \%, all with reasonably good quality. Now, using the PLE result as an additional constraint, we find reasonable fits must have exciton fractions between 50\%  and 58\% for the same data. This is a reasonably tight constraint. However, just because a model is well constrained does not mean it is accurate. Furthermore, the Rabi model gives us no insight into how to modify a sample design in order to change the exciton fraction. We now turn our attention to the shortcomings of the Rabi model, and how to obtain a better model of the polaritons.

The first issue to address is that of linewidth. The equations presented in Section \ref{sec:intro} come from purely real-valued energies. This is equivalent to saying we assumed quantum states with infinite lifetimes or zero linewidth. Real exciton and cavity states have finite lifetimes. In the absence of inhomogeneous broadening due to disorder, this may be included in our Rabi model by simply adding an imaginary piece to the energies. Equation (\ref{eqn:E_polaritons}) does not change, except that the energies are now understood to be complex valued. Equation \ref{eqn:ex_frac_2state} is no longer valid, the most convenient alternative form for the exciton fraction for our purposes is
\begin{equation}\label{eqn:ex_frac_complex_E}
|X_{k_{\parallel}}|^2 = \frac{|E_{up}(k_\parallel)-E_x|^{2}}{g^2 + |E_{up}(k_\parallel)-E_x|^{2}}.
\end{equation}

We can fit Equation (\ref{eqn:E_polaritons}) with complex energies to the extracted dispersion curve from Figure \ref{fig:LP_PL_profile}A, using the results of our PLE measurement to give us an additional constraint. We find that the fit is still under-constrained, even if we use the linewidth of the upper polariton as measured by PLE as a further constraint. There is a family of fits ranging from 35\% excitonic to 58\% excitonic that are all comparably good fits. Fundamentally this is a problem of information. In moving from purely real energies to complex energies we introduced more additional fitting parameters (the imaginary parts), and simply do not have enough measurements available to us to tightly constrain these fits.

Additionally, although complex energies are sufficient for handling homogeneous line broadening, we have reason to believe there is also broadening due to inhomogeneous disorder. Our sample consists of 12 quantum wells all at slightly different locations inside of the microcavity. Differences between these quantum wells would need to be incorporated into Equation (\ref{eqn:starting_hamiltonian}) as additional exciton states. This would be the simplest version of incorporating inhomogeneous disorder. There is almost certainly some number of localized exciton states caused by small defects in the growth of the sample as well.

Furthermore, even if we create a model that takes these issues into account, a question of usefulness arises. We are not only interested in analyzing samples that already exist, but also modifying existing sample designs to produce new samples. For example, we wish to iterate on our designs to produce samples which are close to 50\% excitonic at the flat region in the middle of the sample so as to create polariton structures with minimal energy gradient \cite{yao_dec2022}. Simply using the equations from a two-state model may be sufficient for determining an exciton fraction; however, it tells us nothing about how to modify a design to achieve a desired sample. For these reasons, we pursue an electromagnetic simulation of the sample utilizing the transfer-matrix method (TMM). Utilizing Equation (\ref{eqn:ex_frac_general}), this model will ultimately be able to give us an exciton fraction and be used to design future iterations of a sample design.

Our goal is to produce a working TMM simulation of our samples which we can use to determine exciton fraction of our polaritons. We also wish for our simulation to be accurate even with slight modifications of the sample design so as to predict the polariton characteristics of proposed future samples. For an outline of the TMM equations see Chapter 5 of Ref.~\cite{chuang2009physics}. The central aspect of a TMM simulation is that the sample is treated as a series of infinite slabs of material stacked upon one another. The Fresnel equations are utilized along with some linear algebra, and ultimately with knowledge of the complex refractive index of the materials in the sample, we may efficiently compute the reflection and transmission coefficients of an electromagnetic plane wave incident on the sample. The TMM equations can handle various angles, wavelengths, and polarizations. Thus, a working TMM simulation allows broad reflectivity measurements to be useful constraints on the model of our sample.

We thus want accurate reflectivity data for our samples. In our context, reflectivity $R$ refers to the power reflectivity equal to $|r|^2$, where $r$ is the complex amplitude reflection coefficient. We need to be able to collect the PL data shown in Figure \ref{fig:LP_PL_profile}A, perform the PLE sweep in Figure \ref{fig:PLE_results}D, and collect reflectivity data, at cryogenic temperatures, without having to move the sample or change the optics in a significant way. We performed reflectivity measurements using free standing optics as shown in Figure \ref{fig:optics_setup}. Flipping mirror 1 into the laser's path converted our setup to a reflectivity geometry, same as was used for PL and PLE measurements. Additionally, we had a flip mirror to block the laser and replace it with a broadband white light source. In order to maintain good signal-to-noise ratio throughout our wavelength ranges, a white light source with strong infrared components was preferred. We used the OSL2BIR bulb from Thorlabs. A protected silver mirror from Thorlabs was then placed into the sample plane of the microscope objective, and a series of spectrally resolved and angle-resolved images, which we will call $N$, was taken to establish the input from the white light source to our system. During this collection we also had an aperture in the real-space plane of the setup, which led to crisper images in $k$-space. The silver mirror was then removed and replaced with our cryostat-mounted polariton sample. Care was taken to not change any of the optics in this step. Once our sample was in position, we take a series of images of the sample, which we will call $S$. We also took background images. The background $B_N$ for the $N$ images was obtained by putting nothing at the sample plane of the microscope and letting the beam terminate at infinity. The background $B_S$ for the $S$ images was obtained by putting highly absorbing black optical tape inside the cryostat next to the sample. This allows us to account for most of the background caused by reflection off the cryostat's window. These four measurements were then used to calculate the power reflectivity at all the angles within our numerical aperture and at all the wavelengths in our spectral field of view.

Two corrections were necessary. The reflectivity of our silver mirror $R_M$ is not 100\%, and so we corrected for this using the manufacturer's measured reflectivity (although in practice this is a minimally important correction). Additionally, the cryostat's window was passed through twice, and so the transmission $T_W$ of the cryostat window must be taken into account. This was handled by simply placing the silver mirror inside of our cryostat when we collected the $N$ images.

The measured reflectivity is then
\begin{equation}\label{eqn:refl_calc}
R(\lambda,\theta) = \left( \frac{S(\lambda,\theta) - B_S(\lambda,\theta)}{N(\lambda,\theta) - B_N(\lambda,\theta)} \right) \frac{1}{R_M(\lambda)}.
\end{equation}

\begin{figure}
  \includegraphics[width= 1 \linewidth]{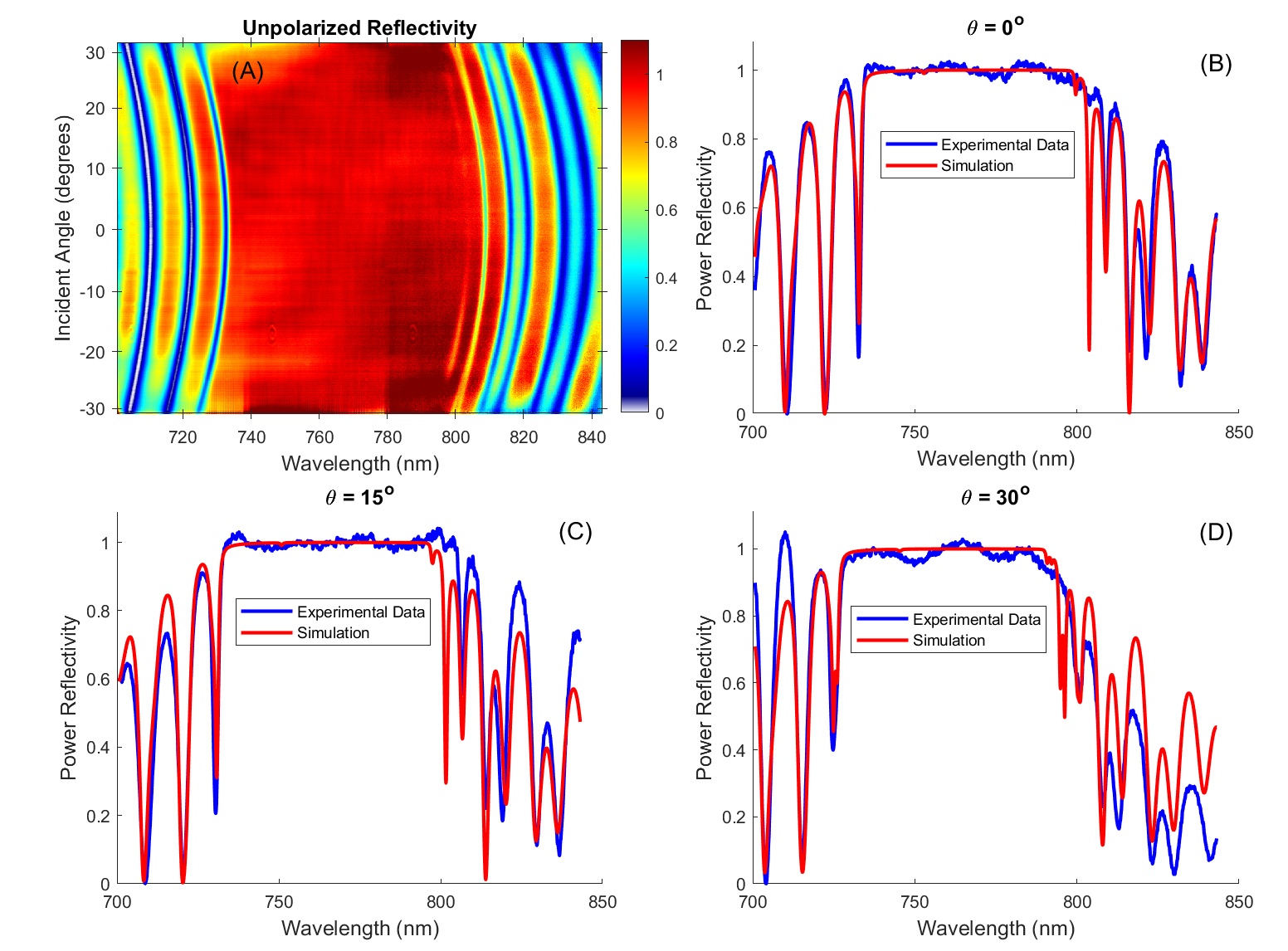}
  \caption{Image A: Reflectivity measurement of our long-lifetime samples. This image is calculated from four measurements as described in equation \ref{eqn:refl_calc}. Images B, C, and D show the agreement between our TMM simulation and the data in image A after our fitting process at three different angles. We see reasonably good agreement. The discrepancy near 805 nm is likely caused by the bulk excitons in the substrate of our sample.}\label{fig:refl_fit}
\end{figure}

This measurement was performed at the same location as the PL and PLE measurements; typical data is shown in Figure \ref{fig:refl_fit}A. Three profiles of the reflectivity along the lines $\theta=0$, $\theta=15$, and $\theta=30$ are included in images \ref{fig:refl_fit}B, \ref{fig:refl_fit}C and \ref{fig:refl_fit}D. We see the expected flat photonic stopband in the center and the Bragg modes clearly on the edges. As discussed above, we can not see either polariton branch in the reflectivity measurements of the long-lifetime samples. We can, however, observe the polariton branches in the short-lifetime samples. The reflectivity measurement of both polariton branches in short-lifetime samples will be utilized in Section \ref{sec:low_Q} to investigate discrepancies between PL,PLE, and reflectivity measurements. 

To create a TMM simulation which matches the data shown in Figure \ref{fig:refl_fit}, we needed data for the index of the main materials in our DBRs; GaAs, AlAs, and the alloy AlGaAs (20\% aluminum) at cryogenic temperatures.  We tried various published data sets and models for our index functions \cite{index_adachi_2,index_aspnes_1,index_Fern_1,index_rakic_1,index_gehrsitz_1,index_papatryfonos_1,index_AlGaAsSb}. Ultimately, we decided to use the index functions given by Reference \onlinecite{index_AlGaAsSb} because they allowed us to tune the temperature and alloy percentage. This is because it is often quite useful to first measure the properties of a structure at room temperature before going to cryogenic temperature. The index functions given in Reference \onlinecite{index_AlGaAsSb} covers all three of our DBR materials over a wide range of temperatures and wavelengths. This particular library, however, does have the downside that the imaginary component of the index functions seem to be consistently larger than what we believe is the the case for our samples, based on the fits of our TMM simulations to our data. Most likely, the discrepancy comes from the difference between high purity MBE grown materials for our structures compared to materials grown through other methods. Higher purity materials will have less impurities, which should reduce absorption, especially in the transparent region below the band gap where absorption is already low, which is the most important spectral region for our model. We found that to have consistency with our data, the imaginary part of the index of refraction of the DBR materials needs to be reduced by about a factor of 100 compared to the values reported in Ref.~\onlinecite{index_AlGaAsSb}, to produce the characteristic flat stopbands shown in Figure \ref{fig:refl_fit}. This results in the imaginary component of the modified index functions of Reference \onlinecite{index_AlGaAsSb} being in reasonable agreement with Reference \onlinecite{index_papatryfonos_1}, which reported index measurements on MBE-grown materials specifically.

 No matter which index reference we use, we will need to tune parameters of our simulation to match our data. We find that simply uniformly scaling the thicknesses of all the layers of the sample in our simulation relative to the designed, nominal values of the widths is not sufficient to bring our simulation into good agreement with our data. We therefore tuned the parameters of our simulation to fit the broadband reflectivity data.  This TMM simulation was coded into a convenient graphical user interface (GUI) based on the Matlab platform, which is available online for download \cite{beaumariage_tmm_matlab}, along with video tutorials of its use. This GUI also includes an optimizer to tune variables in order to align simulation with measurement. Below, we outline what the variables are which we tune in order to produce the good agreement shown in Figures \ref{fig:refl_fit}B, \ref{fig:refl_fit}C, and \ref{fig:refl_fit}D.

The first set of variables we introduced are three scaling factors for the thickness of all GaAs, AlAs, and AlGaAs layers. These variables account for the possibility of systematic overgrowth/undergrowth during the growth process. We tuned these three variables in order to fit our simulation's reflectivity to the measured reflectivity utilizing Matlab's built in least-squares curve fitting function. 
However, in a problem such as this, many local minimums in the $\chi^2$ function exist. Therefore, after running the least-squares function to find a local minimum, we then randomized our variables within 0.2\% of their optimized values to generate the starting conditions for the next run, and ran the least squares function again, in an annealing-type methood. The percentage 0.2\% used here can be termed the ``aggressiveness;'' we found 0.2\% worked best through trial and error. An aggressiveness that is too large will waste a lot of optimization time as the starting conditions for each iteration are often very bad. Similarly, a low aggressiveness will converge towards a global minimum much more slowly.  The algorithm repeats the randomization process a few hundred times, each time randomizing the best minimum found so far.

Varying only the thicknesses of the layers did not give satisfactory fits; the result was a simulation in which the photonic stopband was narrower than we measure. The width of a photonic stopband is directly proportional to the difference between the index of the two DBR materials \cite{kavokin_book}. Therefore, we also allowed the index functions of AlAs and AlGaAs to vary in our fitting procedure. 
We found that good fits could be obtained by simply introducing two scaling factors $\alpha_{AlAs}$ and $\alpha_{AlGaAs}$, and modified our index functions as
\begin{eqnarray}
  n'_{AlAs}(\lambda) &=& \alpha_{AlAs} n_{AlAs}(\lambda) \nonumber \\
  n'_{AlGaAs}(\lambda) &=& \alpha_{AlGaAs} n_{AlGaAs}(\lambda),
\end{eqnarray}
where n is the real part of the index function of the two materials. Similarly, we introduced scaling factors $\beta$ for the imaginary piece of the index functions,
\begin{eqnarray}
  \kappa'_{AlAs}(\lambda) &=& \beta_{AlAs} \kappa_{AlAs}(\lambda) \nonumber \\
  \kappa'_{AlGaAs}(\lambda) &=& \beta_{AlGaAs} \kappa_{AlGaAs}(\lambda),
\end{eqnarray}
where $\kappa$ is the imaginary piece of the index function. Typically we find that the thicknesses of the layers needed to be scaled between $0.95$ and $1.05$ of their designed values, and the $\alpha$ variables need to be scaled between $0.95$ and $1.05$, in order to produce the agreement shown in Figure \ref{fig:refl_fit}. However, as mentioned above, the $\beta$ variables were typically between $0.01$ and $0.001$, likely caused by the high purity of our samples, resulting in low absorption.

Our GUI is capable of fitting multiple reflectivity curves simultaneously. For example, three line profiles at 0, 15, and 30 degrees were taken from the data shown in Figure \ref{fig:refl_fit}A, and are shown in Figures \ref{fig:refl_fit}B, \ref{fig:refl_fit}C, and \ref{fig:refl_fit}D. These three curves are being fit with equal weighting. Our GUI is quite versatile and capable of including in its fits profiles at various angles, temperatures, polarizations, and even other types of data such as ellipsometric and transmission. However, we found that including three profiles at cryogenic temperatures was sufficient for our purposes, and resulted in a reasonably fast optimization time. Standard Matlab measures were taken to optimize our code; letting the simulation run over night is enough time to produce the fits shown in Figure \ref{fig:refl_fit} using a standard modern computer.

We see that tuning the thicknesses of the layers and the index functions results in a simulation which largely agrees with our measured reflectivity over a fairly broad wavelength range. The fit is especially good at normal incidence, except at approximately 805 nm. This region is heavily influenced by the band gap of our GaAs substrate at cryogenic temperatures. Indeed, using a non-resonant laser it is possible to see some PL around 805 nm. When we pump the sample with our white light source, some of the white light is likely absorbed and scatters down to then be re-emitted around 805 nm. This would serve to fill in the dip that should be present in the reflectivity. Our simulation is not capable of accounting for absorption at one wavelength causing emission at another wavelength, and so we believe this explains the discrepancy.

At higher angles, the numerical aperture of our system begins to limit the reliability of our data. Our optics are capable of handling up to $\pm 45$ degrees of light cone, and room temperature measurements of the reflectivity look quite good over this entire range. However, we find that when imaging through our cryostat window it is not possible to collect good reflectivity data beyond 30 degrees. 

We performed these measurements at various locations on our samples. For a rotated MBE growth, all layers should scale by the same percentage as point of observation moves outward radially. When we take the simulation which produces the curves in Figure \ref{fig:refl_fit}, and scale only the thickness of the entire sample uniformly, we find similarly good agreement with experimental reflectivity data measured at other locations on the wafers. This means that we only need to perform the fitting procedure outlined above at one location on the sample, and then we have a working broad reflectivity simulation for all locations of interest on the sample.

\section{\label{sec:TMM_fine} Fitting the TMM Model to PL and PLE}

With a TMM simulation that matches our broader reflectivity measurements, we now turn our attention to the finer details governing the polaritons. Generally, we find these two parts of our fitting to be somewhat independent. The variables we will tune in our simulation in this section have almost no effect on the results shown in Figure \ref{fig:refl_fit}. Similarly, the tuning performed in Section \ref{sec:TMM_broad} has a modest impact on the polaritons we simulate in this section.

\begin{figure*}
    \centering
    \includegraphics[width=1\linewidth]{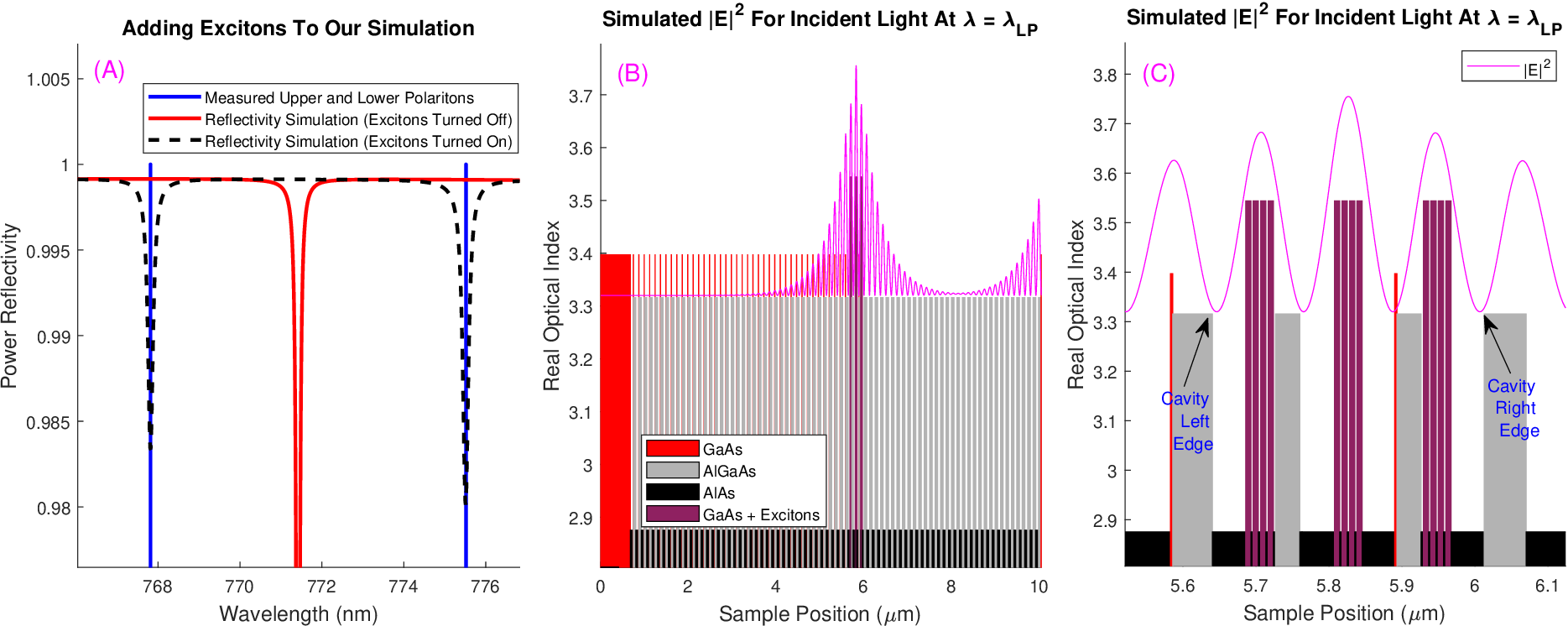}
    \caption{Image A: Our simulation with the excitons turned off, and then with the excitons turned on. We see the cavity mode splits into our upper and lower polariton modes. The vertical blue lines are the experimentally measured values. Image B: The electric field plotted on top of a scale drawing of our sample. We see the electric field confined within the microcavity. Image C: a zoom-in of image B, where we can see the antinodes of the electric field lie on top of the quantum wells, which helps give our samples their strong coupling.}
    \label{fig:turning_on_qws}
\end{figure*}

In Figure \ref{fig:refl_fit}, no traces of the polaritons are seen in the experimental data nor the simulation. This is because the linewidth of the cavity is too narrow for our spectrometer to resolve, and the mesh of the simulation is too coarse to resolve them. However, if we significantly zoom in on our simulation in Figure \ref{fig:refl_fit}B, and increase the density of our simulation mesh, we obtain the red curve in Figure \ref{fig:turning_on_qws}A. Only the bare photon mode is seen because so far, we have ignored the exciton resonance in the quantum wells inside of our microcavity. We must now determine how to include the excitons in our simulation.

Some works accounted for the excitons in the index of refraction of the GaAs quantum wells  \cite{index_gaas_qw}.  We decided, however, to use a simple charged-oscillator model for the excitons \cite{snoke_solidstate_book}. We start with the equation for the complex electronic susceptibility of the excitons,
\begin{equation}\label{eqn:exciton_susceptibility}
\chi_{ex}(E) = A \frac{(E_{ex}^2 - E^2) + i \Gamma E}{(E_{ex}^2 - E^2)^2 + \Gamma^2 E^2},
\end{equation}
where $E_{ex}$ is the natural energy of the exciton, $\Gamma$ describes local damping but is also equal to the full width at half max (FWHM) of the exciton resonance due to homogeneous broadening, and $A$ is an amplitude related the density of oscillators in our quantum well. These three parameters can be tuned in order to bring our simulation into agreement with both our lower polariton PL measurement and our upper polariton PLE measurement.

However, as we will discuss in Section \ref{sec:low_Q}, we also performed this fitting on the data obtained from low-Q samples where we may directly measure both polariton branches through reflectivity. We consistently found that the curvature of the upper polariton of our simulation was larger than what we experimentally measured in reflectivity. That is equivalently, our simulation produced an upper polariton with too small a mass compared to experimental data. We tried a few approaches to fix this. We tried convolving Equation (\ref{eqn:exciton_susceptibility}) with a Gaussian distribution in order to account for inhomogeneous disorder (i.e. a Brendel-Bormann oscillator). We also tried creating a finite series of exciton states to account for the 12 slightly different quantum wells in our sample. We also tried making $E_{ex}$ dependent on angle in order to account for the non-infinite mass of the excitons. However, all these methods required unbelievable numbers in order to work. 

Ultimately, we found the most convincing approach to be a second exciton state at an energy approximately 15-20 meV above the first exciton. This serves to bend the upper polariton downward into agreement with our measurements. This is a reasonable model as we know that there are both heavy-hole and light-hole excitons in the quantum wells, with heavy-hole excitons being the lower energy particle. In this picture with both heavy-hole and light-hole excitons present, our so-called upper polaritons should perhaps be called middle polaritons. However, for consistency, we will continue to call them upper polaritons as we have no need to refer to the third polariton state which lies above the light-hole excitons. 

We add the two excitons into our quantum wells by simply adding their susceptibilities to the susceptibility of the bulk GaAs already in our simulation:
\begin{equation}\label{eqn:susceptibility_addition}
\chi_{QW}(E) = \chi_{GaAs} + \chi_{ex,hh} + \chi_{ex,lh}
\end{equation}

We find we may impose two constraints and still fit our model to our data. We impose that $A_{hh} = A_{lh}$ and that $\Gamma_{hh} = \Gamma_{lh}$. Which means we have 4 parameters governing the excitons in our quantum wells. The two exciton energies $E_{ex,hh}$ and $E_{ex,lh}$, and the amplitude and linewidth they both share $A$ and $\Gamma$.

\begin{figure}
  \includegraphics[width= 1 \linewidth]{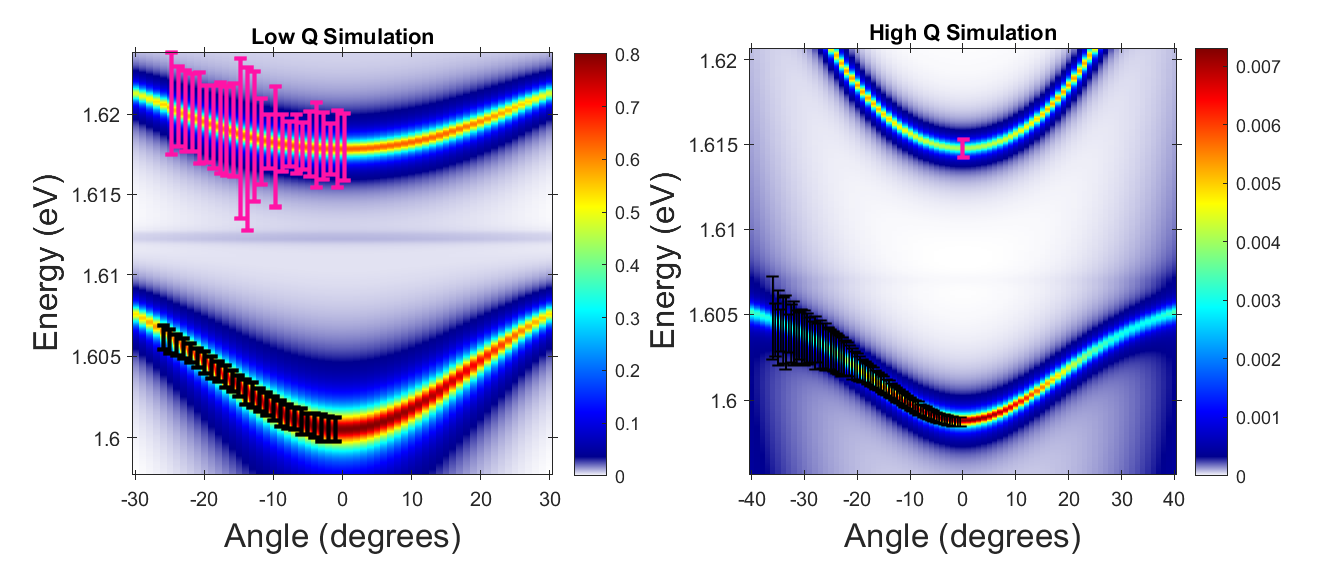}
  \caption{Image A: In the background, our unpolarized inverted reflectivity $1-|r|^2$ from our fitted TMM simulation of the low-Q samples. Plotted on top is half the upper and lower polariton data as seen in reflectivity measurements. Image B: In the background, our unpolarized inverted reflectivity $1-|r|^2$ from our fitted TMM simulation of the high-Q samples with a tight color scale to make the polaritons more visible. Plotted on top of the upper polariton is our PLE measurement. Plotted on top of the lower polariton is half the lower polariton data as measured through photoluminescence via non-resonant pumping. On both images the width of the bars represents the FWHM of the measurements.}
  \label{fig:sim_compare_data}
\end{figure}

When we add just the heavy-hole exciton, we see that the single dip produced in the solid red curve of Figure \ref{fig:turning_on_qws}A splits into two dips as shown in the dashed black curve. When we also include the light-hole exciton, a third dip is present at shorter wavelength outside the field of view. These dips are the polariton branches. Our simulation may be run over a range of angles to create the polariton dispersion curves, as shown in Figure \ref{fig:sim_compare_data}. 

In order to produce good agreement between our simulation and our measured data over a range of angles, we find it necessary to introduce a fifth fitting term at this point. The layers which make up the cavity are illustrated in Figure \ref{fig:turning_on_qws}C. We introduce a scaling term which uniformly scales the thickness of all non-quantum well layers inside the cavity, relative to the rest of the sample. Inside the cavity there are very many thin layers. So, issues such as systematic overgrowth/undergrowth or extra disorder at the interface may be present, and this scaling term allows us to account for such effects.

With these five tunable parameters, we can tune our simulation to match our measurements of the polaritons. We note that our measurements of polaritons in our high-Q samples are not reflectivity measurements, while our simulation is a reflectivity simulation. This means there are two shifts we may need to worry about. These are the difference of the peaks of the lower polariton reflectivity measurements and the PL measurements, and the difference of the peaks of the upper polariton PLE measurements and reflectivity measurements. In Section \ref{sec:low_Q} we show that by good fortune, these two shifts tend to produce errors that offset one another, and so we may proceed forward without worrying about them too much. 

The PL and PLE data are relatively pristine peaks, but our TMM simulation gives dips on top of a background stopband which may have some shape to it. In order to compare the two, at each incident angle of our simulation (each column in Figures \ref{fig:sim_compare_data} A and B) we fit the polariton dips with a Lorentzian line shape subtracted from a background function. Typically the background is a second order polynomial. We perform this at every angle until we have extracted the dispersion curve of the lower polariton $\lambda(\theta)_{lp,sim}$. We similarly perform the same routine on the upper polariton to obtain $\lambda_{up,sim}$, recalling that for high-Q samples we only have to perform this at $\theta = 0$ corresponding to our PLE measurement. 

In order to fit our simulation to our data, we need to define a metric to minimize. First we have the error associated with the lower polariton: 

\begin{equation}\label{eqn:lp_pl_error}
  e_{lp} = \frac{1}{n_{\theta}}\sum_{\theta} |\lambda(\theta)_{lp,sim} - \lambda(\theta)_{lp,pl}|,
\end{equation}

Where $n_{\theta}$ is the number of terms within the summation. Similarly we have the error for the upper polariton: 

\begin{equation}\label{eqn:up_ple_error}
  e_{up} =  |\lambda_{up,sim} - \lambda_{up,PLE}|
\end{equation}

A best fit will involve the minimization of both of these uncertainties; however, we find that sometimes the error of the lower polariton can dominate the upper polariton. So we introduce a weighting factor $\alpha_{up}$:

\begin{equation}\label{eqn:total_error}
  e_{total} = e_{lp} + \alpha_{up} * e_{up}.
\end{equation}

Through trial and error we have found that $\alpha_{up} = 5$ works well. Users who download our code and use it for their own characterization may wish to experiment with this number, which can be easily set in the GUI. Our fitted simulation for our high-Q sample is shown in Figure \ref{fig:sim_compare_data}B. The inverted power reflectivity $1-R$ is plotted in the background image with a tight color scale to make the polariton dips in the stop band visible, and the LP PL data and the PLE data point are plotted on top as bars. The size of the bars corresponds to the FWHM of the data, so the reader may compare the linewidths of our simulation to the linewidths of our data. We see very good agreement overall.

We experimented with alternative forms for Equation (\ref{eqn:lp_pl_error}) with the goal of making the linewidth of our simulation match the linewidth of our experimental data. However, we found this detrimental. Our simulation can not capture several essential factors affecting the linewidths. We know that high-momentum polaritons may scatter down into lower momentum states. Additionally, we know upper polaritons may scatter down into lower polariton states. These effects will result in line broadening \cite{snoke_reanalysis_2023}. Hence, we believe our simulation will always undershoot real linewidths. The $\Gamma$ fitting parameter has almost no effect on the shape of the dispersion curves, and almost exclusively determines the linewidths of our polaritons. So, we remove it from our pool of fitting parameters and fix it to 0.5 meV, which is in good agreement with measurements we performed on bare GaAs quantum well samples. We would expect the linewidth of the lower polariton near $\theta = 0$ to be the least tainted by these scattering effects, and indeed we generally find the linewidth of our simulation to be in modestly good agreement with our PL measurements in that region. 

With our TMM simulation now fitted to our data sets, we may use Equation (\ref{eqn:ex_frac_general}) to extract the exciton fraction via a numerical derivative. We do this by shifting the exciton energies $E_{ex}$ up and down by a small amount $\Delta E_{ex}$, then we extract the corresponding curves of the lower polariton from our simulation, $E_{lp_+}$ and $E_{lp_-}$ respectively. Finally, we can calculate the exciton fraction:
\begin{equation}\label{eqn:ex_frac_numerical}
|X_{k_{\parallel}}|^2 = \frac{d}{dE_x} E_{lp}(k_\parallel) \approx \frac{E_{lp_+}(k_\parallel) - E_{lp_-}(k_\parallel)}{2 \Delta E_{ex}}.
\end{equation}

As $\Delta E_{ex}$ becomes smaller, the simulation mesh needs to become smaller to accurately extract the derivative of the lower polariton curves. We found $\Delta E_{ex} = 0.1$~meV to work well enough. Additionally, because we have two excitons in our simulation, we perform this numerical derivative on both excitons separately to arrive at our heavy-hole and light-hole exciton fractions $|X_{k_\parallel,hh}|^2$ and $|X_{k_\parallel,lh}|^2$. We find the light-hole exciton fraction is generally quite small, which is sensible as the light-hole excitons are further away from the lower polaritons. For the heavy-hole exciton, we see the expected behavior; namely, the lower polaritons are more excitonic at larger angles.

The quantity of interest for most calculations is the total excitonic fraction at a given location on the sample, and at a given in-plane momentum, which we write as
\begin{equation}\label{eqn:total_frac_definition}
|X_{k_{\parallel},total}|^2 = |X_{k_{\parallel},hh}|^2 + |X_{k_{\parallel},lh}|^2.
\end{equation}

\begin{figure}
  \includegraphics[width= 1 \linewidth]{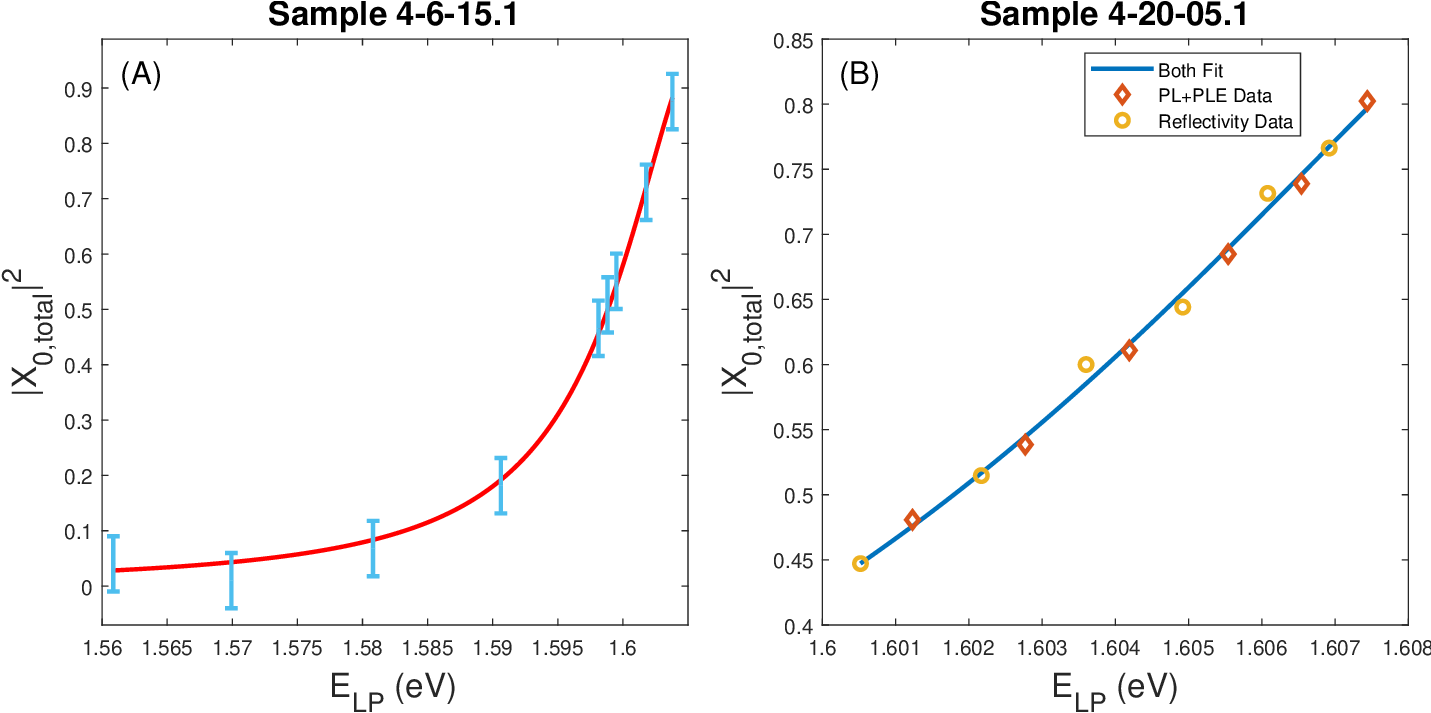}
  \caption{Image A: At nine locations on sample 4-6-15.1 we performed the PL, PLE, and reflectivity measurements described here, and then used our fitted TMM simulation to extract our total exciton fraction at those nine locations. Those 9 data points plotted here are then fit with equation \ref{eqn:ex_frac_2state_fitting}. The values of the two fitting parameters are given in table \ref{tab:exc_fits}.  Image B: We repeated this process on a low-Q sample, and the data is shown and labeled as, ``PL+PLE Data.'' Additionally in the low-Q sample we may measure the upper and lower polariton dispersions directly from the reflectivity measurements at those same locations. We may fit our TMM simulation to those dispersion curves instead, and extract our total exciton fraction, shown here and labeled as ``Reflectivity Data.'' The fit of Equation (\ref{eqn:ex_frac_2state_fitting}) to both sets of data is nearly identical, so we plot a single curve which has been fit to both sets of data together. This combined fit is reported in Table \ref{tab:exc_fits}.}
  \label{fig:hopx_vs_E_lp}
\end{figure}

Specifically, we are typically interested in the total exciton fraction at normal incidence $|X_{0,total}|^2$. We perform the above data collecting and fitting at a range of locations (detunings) on the sample and plot our final result as a function of the lower polartion's energy in Figure \ref{fig:hopx_vs_E_lp}A. This process was also performed on a low-Q sample with the results shown in image B, which is discussed in Section \ref{sec:low_Q}. The error bars are $\pm .05$ for reasons which will be discussed in Section \ref{sec:low_Q}. 

Our characterization problem is now fully solved. By performing a reflectivity measurement, lower polariton PL measurement, and upper polariton PLE measurement we have shown how to extract the total exciton fraction at a single location on the sample, and we have shown this method works over a wide range of detunings. However, not every lab has access to the tools to perform all three measurements; furthermore it is not always prudent to perform all three measurements. For this reason, we seek to create a simple fit of the data so that the exciton fraction can quickly be estimated from just a measurement of the lower polariton's energy. The two-level model from Section \ref{sec:intro} gives us a functional form to try. We start by manipulating equation \ref{eqn:E_polaritons} to solve for $E_c$:
\begin{equation}\label{eqn:E_c_hop_form}
E_c = E_{LP} + \frac{g^2}{E_x - E_{LP}}.
\end{equation}

We are only interested in $k_{\parallel} = 0$ so we drop the $k$-dependence in our notation. We can then plug this into Equation (\ref{eqn:ex_frac_2state}) to arrive at
\begin{equation}\label{eqn:ex_frac_2state_fitting}
|X|^2 = \frac{1}{2} \left( 1 - \frac{ E_{x,global} - E_{LP} - \frac{g_{global}^2}{E_{x,global} - E_{LP}} }{\sqrt{4g_{global}^2 + \big (E_{LP} + \frac{g_{global}^2}{E_{x,global} - E_{LP}} - E_x  \big )^2}} \right).
\end{equation}

We may now view this as a fitting equation for the total exciton fraction at normal incidence as a function of the lower polariton energy, where the exciton energy $E_{x,global}$ and the coupling strength $g_{global}$ are now fitting parameters. This functional form is a two-parameter fit that we found works quite well, as it automatically imposes limiting behavior that should be present in our fit. However, physical meaning should not be inferred from the values of the fitting parameters; for this reason we have attached the ``global'' subscript to help keep this in mind for the reader. We now tune our fitting parameters and arrive at the best fit as shown by the solid line in Figure \ref{fig:hopx_vs_E_lp}. The fit works quite well. The fitted parameter values are given in Table \ref{tab:exc_fits} along with the fitted parameters for many other samples on which we performed similar measurements and fits. (The high-Q data presented thus far is from sample 4-6-15.1). Many of the samples are from our collaborators at Princeton, however our collaborators at Waterloo have also begun producing high-Q samples as well. 

The reader should note that this fit applies to the polariton energy at normal incidence only. At a single location, the polaritons at higher in-plane momentum will also have higher energy, but the exciton fraction at that higher in-plane momentum is not governed by our fit here.

\begin{table*}
\begin{center}
\begin{tabular}{|c|c|c|c|c|c|c|c|}
\hline
\hspace{0cm} Sample \hspace{0cm}  & \hspace{0cm}  $E_{x,global}$ (eV) \hspace{0cm} & \hspace{0cm} $g_{global}$ (meV) \hspace{0cm} & \hspace{0cm} Grower \hspace{0cm}  & \hspace{0cm} Q factor\hspace{0cm} & \hspace{0 cm} \# QWs \hspace{0 cm} \\ \hline
4-6-15.1      & 1.6066    & 7.8051  & Princeton & High & 12   \\ \hline
P8-10-17.1    & 1.6091    & 12.8099 & Princeton & High & 12   \\ \hline
P3-9-18.1     & 1.6134    & 18.8152 & Princeton & High & 12   \\ \hline
P6-19-19.1    & 1.6099    & 9.5462  & Princeton & High & 12   \\ \hline
P9-10-20.2    & 1.6060    & 11.7454 & Princeton & High & 12   \\ \hline
P9-28-20.1    & 1.6089    & 7.7175  & Princeton & High & 12   \\ \hline
P9-29-20.1    & 1.6072    & 12.5079 & Princeton & High & 12   \\ \hline
P1-9-23.1     & 1.6102    & 12.5866 & Princeton & High & 12   \\ \hline
P1-10-23.1    & 1.6138    & 11.6257 & Princeton & High & 12   \\ \hline
4-20-05.1     & 1.6132    & 11.4051 & Princeton & Low  & 12   \\ \hline
10-5-10.1     & 1.6050    & 6.3032  & Princeton & Low  & 3    \\ \hline
G0792         & 1.6196    & 11.5415 & Waterloo  & High & 12   \\ \hline
G0921         & 1.6137    & 10.6059 & Waterloo  & High & 12   \\ \hline
G0985         & 1.5703    & 7.8086  & Waterloo  & High & 12   \\ \hline
\end{tabular}
\end{center}
\caption{The fitting parameters corresponding to equation \ref{eqn:ex_frac_2state_fitting} for some of the many different samples we have used and shared with collaborators over the years. Note that these parameters are just fitting parameters and do not have direct physical meaning. These fits are a convenient form to look up the exciton fraction of a polariton population based only on the k=0 measurement of the lower polariton's energy. These parameters represent the best fits at liquid helium temperatures, nominally 4.2 K. Most of the samples are high-Q factor, consisting of a 40-period bottom DBR and a 32 period top DBR. However, we have a few low-Q samples consisting of a 20 period bottom DBR and a 16 period top DBR. All the samples are 3/2 $\lambda$ cavities, and most have 4 quantum wells per antinode for a total of 12 quantum wells. However, one sample had only one quantum well per antinode for a total of 3 quantum wells. Most of the samples were grown by our collaborators in Princeton, but in the last few years our collaborators at Waterloo have begun growing comparable samples. The high-Q samples are all similar designs, with slight modifications to thicknesses of the various layers.}
\label{tab:exc_fits}
\end{table*}

\section{\label{sec:low_Q} Tests and Validation of our Methods}

\begin{figure}
  \includegraphics[width= 1 \linewidth]{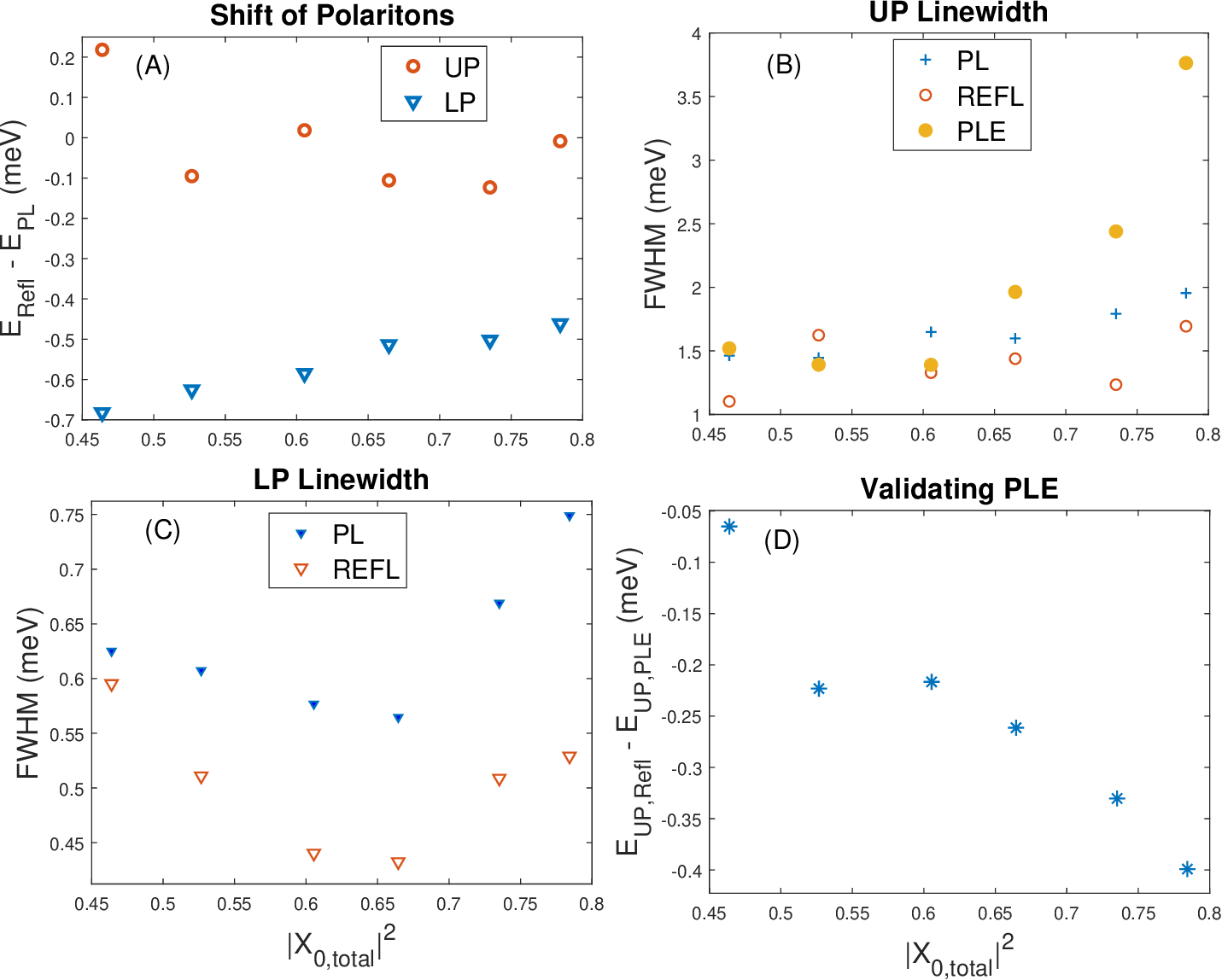}
  \caption{Data taken on short-lifetime samples, which we use to validate our approach in the long-lifetime samples. In all 4 images the horizontal axis is the exciton fraction of the lower polaritons, the reader should take careful note of this. Image A shows the discrepancy between the reflectivity and PL measurements of the polaritons at normal incidence. Image B and C show the linewidths of the upper and lower polaritons through our various measurements. Image D shows the discrepancy between the upper polaritons as measured through reflectivity and PLE.}\label{fig:low_Q_results}
\end{figure}

In this section we test our methods and verify they are sensible. We turn our attention to the short-lifetime sample 4-20-05.1, which was used in prior works of ours (e.g., Ref.~\cite{snoke_science_2007}). We find that PLE measurement of the upper polariton works just as well in low-Q samples as it did in high-Q samples. Three more measurements are possible inside the low-Q samples. Both the upper and lower polariton may be observed as dips inside the stopband when we perform our angle dependent reflectivity measurement as discussed in section \ref{sec:TMM_broad}. For resolving these dips, we swapped to a higher line density grating in our spectrometer to improve our resolution. Additionally, the PL of the upper polariton is also observable when performing the non-resonant pumping measurement as described in Section \ref{sec:ks_pl_refl}. These measurements are significant as they allow us to easily measure the upper polariton over a wide range of angles rather than just normal incidence. Figure \ref{fig:low_Q_results} shows a summary of these measurements performed at 6 locations covering a range of exciton fractions in the low-Q sample. 

One surprising aspect of our measurements is the shift of the polaritons. It is typical in semiconductors to show a discrepancy between PL and reflectivity measurements. Typically the reflectivity measurements are at higher energy. This is the Stokes shift and comes from the disorder and defects of the system. Instead, in the lower polaritons we see a shift in the opposite direction, which is lesser at greater exciton fractions. We think this shift may be be caused by incomplete thermalization of the polaritons, as the more excitonic locations thermalize better. We did a simple power series to confirm our pump laser was not significantly heating up the sample nor introducing a density dependent blue shift. The upper polaritons have an overall smaller shift, some of which is positive. We believe the upper polariton shifts are mainly just exhibiting a scatter in our data, very close to zero. The linewidths are shown in Figures \ref{fig:low_Q_results}B and \ref{fig:low_Q_results}C. We see no clear correlation between the shifts and the linewidths. 

It is a common rule of thumb in the field that the lower polariton linewidth is minimized when the exciton fraction is at 50\%, called the resonant position. However, we find that to be slightly off, and see a minimum closer to 66\%. We have not done a systematic study of this and do not recommend it as a means of finding the resonant position; as discussed above and in Ref.~\cite{snoke_reanalysis_2023}, dynamic scattering of the particles can play a role in linewidth. 

We also may investigate the validity of our PLE measurement. We view the reflectivity result as the more pure measurement of the upper polariton, and so we compare it to our PLE result. The discrepancy between these two measurements is shown in Figure \ref{fig:low_Q_results}D. Note that the horizontal axis is the exciton fraction of the {\em lower} polariton, not the upper polariton. In the Rabi model, the exciton fraction of the upper polariton is  $1 - |X|^2$. This relationship no longer exactly holds, but the two quantities are still inversely related. We see a clear trend in the behavior. When the upper polariton is more excitonic (left side of image), the PLE and reflectivity measurements are in stronger agreement. 

We now use these measurements to estimate the error of our characterization method. To do this, we fit our TMM simulation to the upper and lower polariton curves as measured via reflectivity. One such fit, along with the extracted data is shown in Figure \ref{fig:sim_compare_data}A. As discussed above, the linewidth of the data was not used in the fitting process, as we believe there are significant contributions by factors outside the scope of a purely electromagnetic simulation. Initially, we only had the heavy-hole exciton in our quantum well index function. However, we consistently found that the curvature of the upper polariton was smaller in our data than our simulation was capable of producing, and therefore we could not fit both the upper and lower polariton at the same time. We found we needed to account for the light-hole exciton above the upper polaritons in Equation~(\ref{eqn:susceptibility_addition}) in order to bend the dispersion curve downward.

Once our fitting was done, we used Equations (\ref{eqn:ex_frac_numerical}) and (\ref{eqn:total_frac_definition}) to calculate our total exciton fraction exactly as we did for the high-Q samples above. After doing this at all six locations we are left with the data labeled ``Reflectivity Data'' in Figure \ref{fig:hopx_vs_E_lp}B. We also performed the exact same fits as we did for the high-Q sample, utilizing the PLE result for the upper polariton and the PL curve for the lower polariton in our TMM fitting. The extracted exciton fraction is shown in Figure \ref{fig:hopx_vs_E_lp}B as ``PL+PLE Data''. Note that the twelve total data points were taken at six locations, the different values along the horizontal axis are due to the thermalization shifting witnessed in Figure \ref{fig:low_Q_results}A. 

Surprisingly, we see that both sets of data have nearly the exact same curve when fit by Equation (\ref{eqn:ex_frac_2state_fitting}). When we tried varying the power of the pump laser, this resulted in minimal changes to our data. When we conducted a temperature study using the electronic heater inside our cryostat, we found that as the temperature increases, the entire fitted curve shifts up and to the left. Therefore, temperature variation can not explain why these two measurements result in nearly identical characterization curves. 

Indeed, this is a peculiar result which is still unexplained. It seems as though the shifts in Figure \ref{fig:low_Q_results}A and the discrepancy in Figure \ref{fig:low_Q_results}B perfectly offset one another in our TMM fitting so that the final characterization curve is identical. Although this seems to be a fortuitous coincidence, we may use this data to help us determine error bars for Figure \ref{fig:hopx_vs_E_lp}A. 

The largest discrepancy we see at any one location for the total exciton fraction as measured using either of our two methods is $\Delta |X_{0,total}|^2 = .04$, while the average value was $\Delta |X_{0,total}|^2 = .025$. Additionally, when we account for possible errors in the calibration of our equipment, we think another $\Delta |X_{0,total}|^2 = .02$ is reasonable, so ultimately we settle on the uncertainty of our total exciton fraction to be $\Delta |X_{0,total}|^2 = .05$, which we used for the error bars in Figure \ref{fig:hopx_vs_E_lp}A. 

\begin{figure}
  \includegraphics[width= 1 \linewidth]{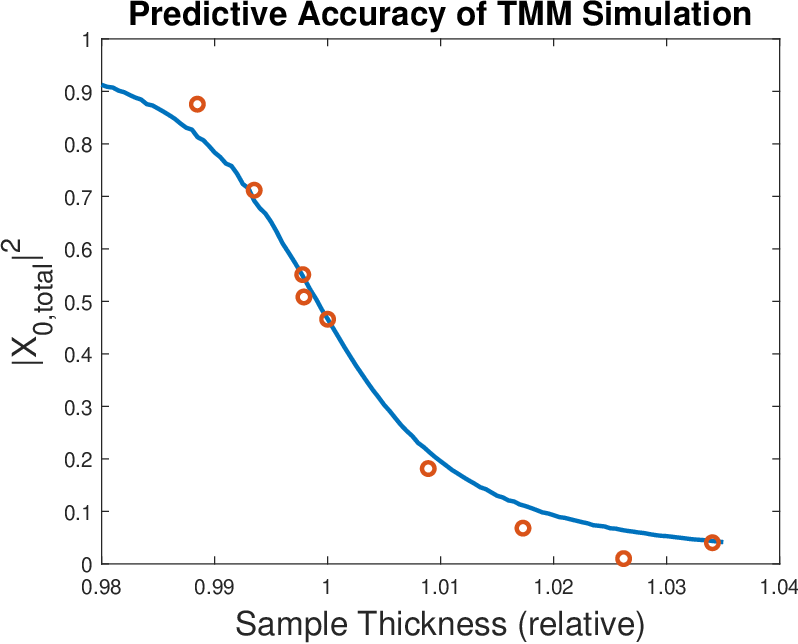}
  \caption{The nine data points are from our full TMM fitting of the reflectivity, PL, and PLE data as described in section \ref{sec:TMM_fine}. The solid line is generated by taking our fitted simulation at the center (denoted as a thickness of one) and uniformly scaling every layer of the simulated sample, and then extracting our exciton fraction. We see that once our simulation has been fit to a location on the sample, it may be used to predict the polaritons at other locations reasonably well. This may also be utilized to predict the polaritons produced by new sample designs.}\label{fig:predictive_TMM}
\end{figure}

Finally, we are interested in the predictive power of our simulation and how it relates to designing new samples. For example, suppose that we have a sample which has no thickness gradient and every location on the sample with a total exciton fraction of $\Delta |X_{0,total}|^2 = .5$, and we wish to modify the design to produce a sample with $\Delta |X_{0,total}|^2 = .7$ by uniformly scaling the layer thicknesses. Can our TMM simulation accurately tell us which changes are necessary to produce this outcome? In order to test this, we first fit our TMM simulation to the central location of the data set in Figure \ref{fig:hopx_vs_E_lp}A, to which we assign a relative sample thickness of one. We can then uniformly scale every layer of the simulated sample by a factor ranging from 0.98 to 1.035 and extracted the exciton fraction from our simulation to produce the solid line shown in Figure \ref{fig:predictive_TMM}. We compare this line to the true result we got from our full TMM fitting as described in section \ref{sec:TMM_fine}. We see that the predictive accuracy is fairly good when scaling by a few percent. Our simulation may be used to predict how small changes will affect future samples.  

\section{\label{sec:results} Conclusions}
We have shown how to measure the upper polariton in high-Q samples using PLE, as well as how to calibrate an angle resolved optical system for PL and reflectivity measurements. These three measurements are fit by our TMM simulation in order to arrive at a total exciton fraction. These methods were also tested on a low-Q sample by fitting the dispersion curves as measured by reflectivity, we discovered the light-hole exciton plays an important role in the curvature of the upper polariton at high angle.

The software we created to do this fitting is controlled easily with a GUI and available for free download online along with several videos demonstrating how to use it. This process was repeated at multiple locations on the sample and ultimately a two parameter fit is created to relate lower polariton energy to exciton fraction. This entire procedure was performed on over a dozen samples including low-Q samples with both 12 quantum wells and 3 quantum wells. Additionally, this process was tested on samples grown by both our collaborators at Princeton as well as our collaborators at Waterloo. 

While more complicated than a simple Rabi model, our TMM fitting procedure has the advantage that once it has been fit to an existing sample, it may predict the affects of small changes to the design.

\begin{acknowledgments}
This work has been supported by the National Science Foundation through Grant No.  DMR-2306977. 

Author Contributions: J.B. carried out most of the sample design, experimental measurements, software writing, data processing, and writing of the paper. Z.S. helped collect the data on sample 4-6-15.1 and helped develop the optical system and methods. H.A. collected and processed the data on samples P1-10-23.1 and P1-9-23.1. Q.Y. collected and processed the data on sample G0985. D.M.M. helped develop the angle calibration methods. M.S. helped develop the PLE measurement method and designed several of the samples. L.P., K.W., and K.B. grew the samples from Princeton, and Z.R.W. and M.C.A.T. grew the samples from Waterloo. D.W.S guided and oversaw all the data measurements, processing, and writing. 

\end{acknowledgments}

%

\end{document}